\documentclass{article}

\PassOptionsToPackage{numbers, compress}{natbib}

\usepackage[preprint]{neurips_2021}

\usepackage{natbib}
\usepackage[utf8]{inputenc} 
\usepackage[T1]{fontenc}    
\usepackage{hyperref}       
\usepackage{url}            
\usepackage{booktabs}       
\usepackage{amsfonts}       
\usepackage{nicefrac}       
\usepackage{microtype}      
\usepackage{xcolor}         
\usepackage{multirow}

\title{Benchmarks and leaderboards for sound demixing tasks}

\author{
  Roman Solovyev\\
  Institute for Design Problems in Microelectronics of Russian Academy of Sciences \\
  Moscow, Russian Federation \\
  \texttt{roman.solovyev.zf@gmail.com} \\
  \And
  Alexander Stempkovskiy\\
  Institute for Design Problems in Microelectronics of Russian Academy of Sciences \\
  Moscow, Russian Federation \\
  \texttt{stemp@ippm.ru} \\
  \And
  Tatiana Habruseva\\
  Independent researcher \\
  Cork, Ireland \\
   \texttt{tatigabru@gmail.com} \\
}

\begin{document}

\maketitle

\begin{abstract}
Music demixing is the task of separating different tracks from the given single audio signal into components, such as drums, bass, and vocals from the rest of the accompaniment. Separation of sources is useful for a range of areas, including entertainment and hearing aids. 

In this paper, we introduce two new benchmarks for the sound source separation tasks and compare popular models for sound demixing, as well as their ensembles, on these benchmarks. For the models' assessments, we provide the leaderboard at \url{https://mvsep.com/quality\_checker/}, giving a comparison for a range of models. The new benchmark datasets are available for download. 

We also develop a novel approach for audio separation, based on the ensembling of different models that are suited best for the particular stem. The proposed solution was evaluated in the context of the Music Demixing Challenge 2023 and achieved top results in different tracks of the challenge. The code and the approach are open-sourced on GitHub.
\end{abstract}

\section{Introduction}
Music demixing is the task of separating audio signals (a mixture) into individual components. Traditionally, music demixing challenges and benchmarks consider the separation of audio into four stems: drums, bass, vocals, and other (all the other instruments). 

Sound separation technology can be employed in different areas, ranging from entertainment to hearing aids. For example, it is used for reviving the sound of classic movies~\cite{sony} where the original soundtrack contains dialogue, music, and sound effects mixed in mono or stereo tracks. Music separation technology can be used for cleaning up human voices recorded through microphones from the surrounding noise and help voice recognition, cleaning vocal sounds in hearing aids and during phone calls. Vocal suppression in songs can enhance karaoke experiences by allowing people to sing together with the original song as the background (where the vocals were suppressed) so that users do not have to pick from a limited set of covers specifically produced for karaoke~\cite{karaoke}. The music separation also appeals to professional creators by allowing unprecedented remixing of songs, surpassing traditional methods like equalizers. 

There is a wide range of deep learning architectures for audio source separation. Most of them follow the encoder-decoder architecture inspired by U-net architecture~\cite{Ronneberger2015}. The encoder extracts representative feature maps of different spatial dimensions, and the decoder reconstructs a full-resolution semantic mask. Different approaches have been used for music demixing, based both on waveform~\cite{Defossez19,waveunet} and spectrogram domains~\cite{defossez2021hybrid}. The most popular models up to date include Spleeter~\cite{Spleeter}, Unmix, Demucs (version 2, 3 and 4)~\cite{rouard2022hybrid,defossez2021hybrid}, MDX-Net~\cite{Kim21} and ultimate vocal remover (UVR)~\cite{uvr}.

To evaluate audio separation approaches, various benchmarks, and challenges were developed. In most of the challenges, participants were tasked to separate a song into four stems: vocals, bass, drums, and other (signals of all instruments and other accompaniment). The popular datasets include MUSDB18~\cite{musdb18} and MUSDB18-HQ~\cite{musdb18-hq} used in SiSEC MUS challenge~\cite{SiSEC18} and AICrowd Music Demixing Challenge 2021 (MDX21)~\cite{MDX21}, Divide and Remaster~\cite{DnR} used in the recent Sound Demixing Challenge 2023 (SDX23)~\cite{SDX23}, AudioSet~\cite{AudioSet}. The AudioSet has weakly-labeled data, Divide and Remaster is the synthetic dataset, and the performance metrics obtained on it do not often reproduce on the real data. The popularity of these datasets leads to the community overfitting them. Alternative benchmarks are useful for assessing the generalization of the algorithms and comparing them on the independent leaderboards. 

Here, we introduce two new benchmarks for the sound demixing tasks and provide detailed leaderboards to compare popular models and their ensembles. The full leaderboards can be found at \url{https://mvsep.com/quality\_checker/} for both datasets. The website allows testing any models by uploading predictions, so the leaderboards are dynamic. Using our findings, we developed an ensemble approach to the recent Sound Demixing Challenge 2023 (SDX2023) and achieved top results in different tracks of the competition.

We present the solution to the SDX2023 challenge, hosted at \url{https://www.aicrowd.com}. Our algorithm filters vocals before separating other stems and employs weighted ensembling of different models and checkpoints. The source code is publicly available at GitHub~\cite{zfturbo}.

\section{Overview of the popular sound source separation models}

In this section, we provide a brief overview of the models used in the SDX23 and our ablation study. The detail for each model architecture can be found in the corresponding referenced papers and/or codebase. 
\begin{itemize}
\item Demucs is based on a U-Net convolutional architecture inspired by Wave-U-Net~\cite{waveunet} with the innermost layers replaced by a cross-domain Transformer Encoder. The model is capable of separating drums, bass, and vocals from the rest of the accompaniment. From the initial publication, different versions of the Demux models evolved, with v4 being the latest at the time of the competition. The v4 version features Hybrid Transformer Demucs~\cite{rouard2022hybrid,defossez2021hybrid}, a hybrid spectrogram/waveform separation model with the innermost layers replaced by a cross-domain Transformer Encoder. The Transformer uses self-attention~\cite{attention} within each domain and cross-attention across domains. 
\item MDX-Net model~\cite{Kim21}, code: \url{https://github.com/kuielab/mdx-net}, is a two-stream neural network for music demixing, KUIELab-MDX-Net. The model has a time-frequency branch and a time-domain branch and blends results from two streams to generate the final estimation. The model got the top places in the MDX21 competition. MDX-Net consists of six networks, all trained separately, see details in~\cite{Kim21}. 
\item Ultimate vocal remover (UVR)~\cite{uvr} is a library with a range of pre-trained models for vocals separation: \url{https://github.com/Anjok07/ultimatevocalremovergui}, which includes pre-trained MDX-Net and retrained Demucs models described above, and it’s own architecture models for vocal separation.

\item Spleeter~\cite{Spleeter}, code: \url{https://github.com/deezer/spleeter}, is an open-sourced library/tool for music source separation containing pre-trained models. It includes pre-trained models for vocals separation, classing four stems separation (vocals, bass, drums, and other) and 5 stems separation with an extra piano stem. The pre-trained models are 12-layer U-nets~\cite{Pretet2019} with the encoder-decoder CNN architecture with skip connections (6 layers for the encoder and 6 for the decoder). It was one of the first successful neural network models for music source separation and was released by the deezer company team.
\item Open-Unmix is a library with deep neural networks for music source separations into classical four stems. The core of Open-Unmix is a three-layer bidirectional LSTM network~\cite{Lstm1997}. Due to its recurrent nature, the model can be trained and evaluated on an arbitrary length of audio signals. The models are pre-trained on the freely available MUSDB18 dataset. This model also has an UmXL version trained on additional data with Non-commercial usage restrictions. Code: \url{https://github.com/sigsep/open-unmix-pytorch}.

\item Zero-shot Audio Source Separation~\cite{Chen21}, code: \url{https://github.com/RetroCirce/Zero\_Shot\_Audio\_Source\_Separation}, is a three-components pipeline for universal audio source separator train on a large weakly-labeled dataset, AudioSet~\cite{AudioSet}. 
\item Danna Sep, code: \url{https://github.com/yoyololicon/danna-sep}, is one of the top models of the MDX21 Challenge~\cite{MDX21}. This algorithm took 3rd place on Leaderboard A in the MDX21 challenge. 
\item Byte Dance, code: \url{https://github.com/bytedance/music\_source\_separation} is an open-sourced music source separation library based on ResUNet~\cite{kong2021decoupling}.
The algorithm took second place in the vocals category on Leaderboard A at MDX21. It was trained on MUSDB18-HQ data.
\item Band-Split RNN~\cite{band-split-rnn}, a non-official code implementation: \url{https://github.com/amanteur/BandSplitRNN-Pytorch}, is a highly scoring model on the MUSDB18 dataset leaderboard.
\end{itemize}

\section{Evaluation}
\label{sec:eval}
The signal-to-distortion ratio (SDR) is a common metric used to evaluate the quality of audio separation algorithms. It is a measure of how well the desired audio sources have been separated from the mixture, while minimizing the distortion caused by residual interference.
The SDR is defined as follows:
\begin{equation}
SDR_{stem} = 10 \cdot \log_{10} \left( \frac{\sum\limits_{n=1}^{N} s_{stem,n}^2}{\sum\limits_{n=1}^{N} e_{stem,n}^2} \right)
\end{equation}
where $s_{stem,n}$ is the waveform of the ground truth and $e_{stem,n}$ denotes the waveform of the estimate. The higher the SDR score, the better the output of the system is.

To rank the entire system, the average SDR across all stems is used for each record:
\begin{equation}
SDR_{record} = \frac{1}{N} \sum_{i=1}^{N} SDR_i ,
\end{equation}
where $N$ is the total number of stems in a given record, and $SDR_i$ is the SDR value for the $i$-th stem.

Finally, the overall score $SDR_{total}$ is the average of the scores for all records in the test set. 

The SDR metrics are widely used to compare different sound demixing models in challenges, and we will use these metrics in the leaderboards and ablation study reported in this work.
\section{The synthetic dataset for vocal separation, Synth MVSep 
}
The synthetic dataset for the vocal separation task was created by combining random vocal and instrumental samples, publicly available on the internet. The sourced samples were separated into two sets (vocal-only and instrumental-only) and then randomly mixed together. The mixtures may not always sound like a real melody, but they allow for testing audio separation methods. 

Synth MVSep dataset consists of $100$ tracks, each with a duration of exactly one minute and a sample rate of $44.1$ kHz. The data is available for downloading at \url{https://mvsep.com/quality\_checker/}; the unzipped data size is around $1.9$ GB. 

Synth MVSep also contains audio files with separate instrumental and vocal parts of the mixture, used for the blind testing of audio separation algorithms on the server. This data is assessed through the evaluation server at https://mvsep.com/quality\_checker/ and is not open for direct downloading to provide a fair comparison for all participants. 

\begin{table}[htb]
\caption{Comparison of various single models based on SDR scores for instrumental and vocal separation on the Synth MVSep dataset}
  \centering
  \begin{tabular}{p{9.8cm}cc}
  \toprule
  Model and description &SDR instr&SDR vocal\\ 
  \midrule
    Our MDX23 (ZFTurbo) model & 11.11 & 11.40\\
    MDX-Net model, trained by Kimberley Jensen. Predict the vocals part (FFT: 7680) & 10.83 & 11.15\\
    MDX-Net: UVR-MDX-NET Inst 3. MDX model, trained by UVR team. Predict instrumental part (FFT: 6144) & 10.61 & 11.01\\
    MDX-B: the best vocals model from MDX21 competition& 10.45 & 10.88\\
    Demucs4 $|$ htdemucs\_ft. Demucs4 HT algorithm trained on 800 songs. Fine-tuned for each stem independently. Shifts: 1, overlap 0.75 & 9.96 & 10.26\\
    Demucs4 HT algorithm trained on 800 songs. Fine-tuned for each stem independently. Default from repository& 9.94 & 10.23\\
    Demucs3, winning algorithm from MDX21 competition& 9.48 & 9.78\\
    UVR initial vocal model (Agg: 0.9)& 9.12 & 9.44\\
    Danna Sep: \url{https://github.com/yoyololicon/danna-sep}& 8.30 & 8.59\\
    Demucs2 algorithm & 8.24 & 8.53\\
    Unmix XL: the best Open-Unmix model from: \url{https://github.com/sigsep/open-unmix-pytorch}& 8.16 & 8.45\\
    Byte Dance library & 7.69 & 7.98\\
    Spleeter algorithm (4 stems): \url{https://github.com/deezer/spleeter} & 7.02 & 7.31\\
  \bottomrule
  \end{tabular}
  \label{tab:models}
\end{table}

The Synth MVSep dataset benchmark was used to evaluate and compare single models and their ensembles. Table~\ref{tab:models} shows examples of single models assessed and their performance metrics. Here we list the popular models described in section 2. The full leaderboard for single models and their ensembles can be found at \url{https://mvsep.com/quality\_checker/}.

\begin{table}[htb]
    \centering
    \caption{Selected ensembles of models and their SDR metrics on the Synth MVSep dataset}
    \label{tab:ensembles}
    \begin{tabular}{p{7cm}ccp{3cm}}
        \toprule
        Models in the ensemble & SDR instr & SDR vocal & Description \\
        \midrule
        MDX-Net: UVR-MDX-NET Inst 3 + UVR-MDX-NET Inst Main + Kim\_vocal\_1 + Demucs: v4 $|$ htdemucs\_ft - Shifts: 2 & 11.26 & 11.61 & 3 MDX + 1 Demucs4 \\
        MDX-Net: UVR-MDX-NET Inst 3 + kim vocal model + Demucs: v4 $|$ htdemucs\_ft - Shifts: 2, Avg & 11.22 & 11.54 & 
        2 MDX + 1 Demucs4\\
        MDX-B(ONNX) + MVSep Vocal + Demucs4 HT. Weights: [3, 2, 2] & 10.99 & 11.29 & 1 MDX + 1 Demucs4 + 1 ByteDance\\
        UVR-MDX-NET Inst 3 and htdemucs\_ft & 10.88 & 11.20 & 1 MDX + 1 Demucs4 \\
        \bottomrule
    \end{tabular}
\end{table}

As seen from Table~\ref{tab:models}, the top-performing models are variations of the MDX algorithm from the open-source project Ultimate Vocal Remover, UVR-MDX~\cite{uvr}. The algorithms differ in the training data used and the length of the fast Fourier transform (FFT). For the top-performing model, the FFT had a length of $7680$, while the others had a length of $6144$.

Ensembles of models are widely used when real-time inference is not required. Combining predictions from different models usually generalizes better and gives more accurate results compared to single models~\cite{okun2011ensembles}. Table~\ref{tab:ensembles} shows results obtained on the Synth MVSep benchmark for the ensembles. As one might expect, the ensembles outperform single models by a margin, but at a considerable computational cost. The averaging of more models also tends to have better performance.

\section{The multisong dataset for music separation, Multisong MVSep 
}
The Multsong MVSep dataset is composed of $100$ publicly available compositions from a variety of genres. The dataset consists of 100 tracks, each exactly one minute long with a sample rate of $44.1$ kHz. The data is available at \url{https://mvsep.com/quality\_checker/}; the unzipped size is around $1.8$ GB. 

The genres in the Multisong MVSep include Acoustic, Folk, Modern Blues, American Roots Rock, Modern Country, Ambient, Beats, Dance, Deep House, Disco, Drum n Bass, Electro, Euro Pop, Future Bass, House, Soft House, Funk, Alternative Hip Hop, Mainstream Hip Hop, Old School Hip Hop, Trap, Acid Jazz, Big Band, Modern Jazz, Smooth Jazz, Bossa Nova, Modern Latin, Salsa, 1970s Pop, 1980s Pop, 1990s Pop, 2000s Pop, 2010s Pop, 2020s Pop, Afrobeats, Indie Pop, K-pop, Synth Pop, RnB, Soul, 1960s Rock, Alternative, Hard Rock, Punk, Modern Hymns, Praise \& Worship, and India. Due to such genre diversity, models that perform well on this benchmark generalize well and are universal. There is a little chance that some of the melodies in the test Multisong MVSep were occasionally used to train some of the models, creating a data leak. The dataset also contains audio files with separate four parts of the mixtures: vocals, bass, drums, and other accompaniment, that make up the compositions. This part is closed from direct downloading for fair testing of algorithms on our evaluation server. 

\begin{table}[htb]
  \centering
  \caption{Selected models for music separation and their performance metrics on the Multisong MVSep dataset (only vocals and instrumental).}
  \label{table:models}
  \begin{tabular}{p{9.8cm}cc}
\toprule
Model and description&SDR instr&SDR vocal\\ 
\midrule
\textbf{MDX23 (ZFTurbo) contest model} & \textbf{15.82} & \textbf{9.56}\\ 
MDX | Kim\_Vocal\_1: MDX model, trained by Kimberley Jensen. Predict vocals part (FFT: 7680)& 15.70 & 9.36\\ 
MDX | UVR-MDX-NET\_Main\_427: MDX model, trained by UVR team. Predict vocals part (FFT: 6144)& 15.61 & 9.32\\ 
MDX: UVR-MDX-NET Inst HQ 2: MDX model, trained by UVR team. Predict instrumental part (FFT: 6144)& 15.22 & 9.11\\
Demux4 HT (demucs\_ft, shifts=1, overlap=0.95) algorithm trained on 800 songs. Fine-tuned for each stem & 14.73 & 8.42\\
Demucs3 (demucs\_mmi) algorithm trained on 800 songs & 14.54 & 8.24\\ 
Demucs4 HT (htdemucs) algorithm trained on 800 songs. Tuned for average SDR for 4 stems & 14.49 & 8.18\\
Demucs4 HT (htdemucs\_6s) algorithm, which includes 6 stems (additional piano and guitar) & 14.48 & 8.17\\
Demucs3 Model B, winning algorithm from MDX21 competition& 14.44 & 8.13\\
UVR vocal models, 0\_HP2-4BAND-3090\_4band\_arch-500m\_1 & 13.59 & 7.28\\
Unmix XL, the best Open-Unmix model  & 12.97 & 6.66\\
Demucs2 algorithm & 12.90 & 6.60\\ 
Spleeter algorithm (2 stems) & 12.05 & 5.81\\ 
\bottomrule
\end{tabular}
\end{table}

\begin{table}[htb]
\centering
\caption{Selected models for music separation and their performance metrics on the Multisong MVSep dataset (only bass, drums, and other).}
\label{table:models2}
\begin{tabular}{p{8.2cm}ccc}
\toprule
Model and description &SDR bass&SDR drums&SDR other \\ 
\midrule
\textbf{MDX23 (ZFTurbo) contest model} & \textbf{12.50} & \textbf{11.70} & \textbf{6.63}\\ 
Demux4 HT (htdemucs\_ft, shifts=10, overlap=0.95) algorithm trained on 800 songs. Fine-tuned for each stem independently & 12.24 & 11.41 & 5.84\\
Demux4 HT (htdemucs\_ft, shifts=1, overlap=0.95) algorithm, fine-tuned for each stem independently; 10 times faster than the previous & 12.24 & 11.40 & 5.84\\ 
Demucs4 HT (htdemucs\_ft, shifts=1, overlap=0.25). Default variant from repository& 12.05 & 11.24 & 5.74\\ 
Demucs4 HT (htdemucs) algorithm trained on 800 songs. Tuned for average SDR for 4 stems & 11.74 & 10.90 & 5.57\\ 
Demucs4 HT (htdemucs\_6s) algorithm, which includes 6 stems (additional piano and guitar) & 11.42 & 10.59 & —\\ 
Demux3 (hdemucs\_mmi, shifts=4, overlap=0.75) algorithm trained on 800 songs& 11.25 & 10.76 & 5.50\\
Demucs3 Model B, winning algorithm from MDX21 competition& 10.69 & 10.27 & 5.35\\
Demucs2 algorithm& 9.01 & 8.24 & 3.84\\
Unmix XL, the best Open-Unmix model & 8.45 & 7.92 & 4.41\\
Spleeter algorithm (4 stems) & 7.32 & 6.89 & 3.49\\
\bottomrule
\end{tabular}
\end{table}

There are separately-ranked leaderboards for the four stems: bass, drums, vocals, and other, and for the all-instrumental (non-vocal) together:
\begin{itemize}
\item bass: \url{https://mvsep.com/quality\_checker/leaderboard2.php?sort=bass}
\item drums: \url{https://mvsep.com/quality\_checker/leaderboard2.php?sort=drums}
\item other: \url{https://mvsep.com/quality\_checker/leaderboard2.php?sort=other}
\item vocals: \url{https://mvsep.com/quality\_checker/leaderboard2.php?sort=vocals}
\item instrumental: \url{https://mvsep.com/quality_checker/leaderboard2.php?sort=instrum}
\end{itemize}

Tables~\ref{table:models} and \ref{table:models2} summarize the results for the selected models assessed and their SDR performance metrics. The full leaderboards are available at the links above.

It can be seen from Table~\ref{table:models} and Table~\ref{table:models2} that various Demucs4 HT models variants dominate all stems except for vocals and instruments separation. Models based on the MDX algorithms are the best for separating the vocals. Therefore, ensembles of different models used for vocal and non-vocal stems are expected to provide the top overall performance.

The ensemble approach was evaluated on the recent Sound demixing challenge 2023. The challenge, approach, and methodology are discussed in the following chapters.

\section{The Sound demixing challenge 2023 (SDX23)}
The challenge included two tracks: music demixing (MDX) and cinematic sound demixing (CDX). The music demixing track is a challenge focused on music source separation. Given an audio signal as input (a “mixture”), participants were tasked to decompose it into four tracks: “vocals”, “bass”, “drums”, and “others” (all other instruments). In the cinematic sound demixing task, participants were tasked with separating the audio of a movie into three tracks: dialogue, sound effects, and music. This task has many applications, ranging from language dubbing to up-mixing of old movies to spatial audio and user interfaces for flexible listening. The challenge was organized by five companies: Sony, Moises.AI, Mitsubishi Electric Research Labs, AudioShake, and Meta~\cite{MDX21,SDX23}.

\subsection{The Cinematic sound demixing track}
The challenge had different leaderboards, depending on the data that could be used for training. For cinematic sound demixing track, Leaderboard A imposed constraints on the training data allowing only usage of the challenge "Divide-and-Remaster" datasets, while leaderboard B allowed any training data. 

\subsubsection{Divide and Remaster dataset}
Divide and Remaster (DnR)~\cite{DnR} is a synthetic dataset created to train and test the separation of a monaural audio signal into speech, music, and sound effects/background stems. The dataset is composed of artificial mixtures using audio from the librispeech, free music archive, and Freesound Dataset 50k~\cite{DnR}. Each mixture in the DnR dataset is 60 seconds long with the sources not fully overlapped. The data is split into training ($3295$ mixtures), validation ($440$ mixtures), and testing ($652$ mixtures) subsets. The audio mixtures are encoded as 16-bit .wav files at a sampling rate of $44.1$ kHz. There are four files for each mixture containing the mixture itself and separately: music, speech, and sound effects wav files. The dataset also includes metadata for the original audio used to compose the mixture (transcriptions for speech, sound classes for sound effects, and genre labels for music). Details about the data generation process can be found at the following link: \url{https://github.com/darius522/dnr-util}.

\subsubsection{The Cinematic sound demixing track, Leaderboard B solution}
The analysis of the models and their ensembles on the leaderboards discussed above allows for choosing the best models for different stems. The main idea here is to separate the vocals first, using a very high-quality model, and then apply a model trained on the DnR dataset to the remaining part (music and effects). 

For the offline validation, the organizers provided two tracks from the test set. The metrics on these tracks, however, were not strongly correlated with the leaderboard, presumably due to the tiny size of the validation set.

To separate the vocals, we used a combination of three pre-trained models: UVR-MDX1 (checkpoint: Kim\_Vocal\_1.onnx \cite{kimvocal}), UVR-MDX2 (checkpoint: UVR--MDX--NET--Inst\_HQ\_2.onnx \cite{uvrmodelhq2}) from Ultimate Vocal Remover project \cite{uvr}, and Demucs\_ft (vocal-only model) from Demucs4 library \cite{demucs4}. The vocals were separated independently by all of these models, then the results were combined with weights. The models' ensemble is summarized below:
\begin{itemize}
\item $vocals_1$ = UVR-MDX1(mixture, overlap=0.6),
\item $vocals_2$ = UVR-MDX2(mixture, overlap=0.6),
\item $vocals_3$ = Demucs4(mixture, 'demucs\_ft', shifts=1, overlap=0.6),
\end{itemize}

\begin{equation}
vocals = \frac{\sum_{i=1} w_i vocals_i}{\sum_{i=1} w_i}
\end{equation}

We tried different weights for the vocals' ensemble; weights $10$, $4$, and $2$ for UVR-MDX1, UVR-MDX2, and Demucs4, respectively, produced optimal results. 

\begin{table}[htb]
\caption{Ablation study for the Cinematic sound demixing track solution}
\label{table:CDX}
\centering
\begin{tabular}{lcccc}
\toprule
Model & SDR dialog & SDR SFX & SDR music & Mean SDR\\
\midrule
CDX23 best ensemble model, val1 & 14.927 & 3.780 & 8.060 & 8.922\\
CDX23 best ensemble model, val2 & 9.949 & 6.377 & 6.429 & 7.585\\
CDX23 best ensemble model, public LB & 15.056 & 3.954 & 4.069 & 7.693\\
Demucs4 (single), val1 & 13.887 & 2.781 & 2.494 & 6.387\\
Demucs4 (single), val2 & 14.151 & 7.740 & 7.012 & 9.634\\
Demucs4 (single), leaderboard & 6.774 & 0.984 & 0.296 & 2.685\\
\bottomrule
\end{tabular}
\end{table}

After obtaining high-quality vocals part we can subtract it from the original track to obtain the instrumental part. To separate the instrumental into two remaining stems, we trained two versions of the Demucs4 model \cite{demucs4} on the DnR dataset. The first Demucs4 model was trained using the standard protocol for all 3 stems, while the second Demucs4 model was trained only on two stems: SFX and music, excluding vocals. We used several checkpoints of each of the models to average predictions and obtain better generalization. Therefore, the final inference was run ten times in total.

Our Demucs4 model itself showed promising results on the validation set (see Table~\ref{table:CDX}), however, the metrics did not fully correlate with the leaderboard, and the model performance was poor on the public test set. Table~\ref{table:CDX} shows the results of the ablation study with and without a separate vocal removal. To We used two different validation sets: $val1$ - validation on two tracks provided by organizers, and $val2$ - a subset of 20 random tracks from DnR test set. 

During the competition, we note that DnR dataset contains vocals in the music part sometimes. Our SDR for vocals on the leaderboard is very high, but our vocals model extracts all vocals from audio. Based on this, we made a conclusion that "music" in the competition dataset most likely never contains vocals.

The final ensemble achieved one of the top results in the challenge, leaderboard B. The final results will be updated when the competition ends.

\subsection{The Music Demixing Track}
The task for the Music Demixing track (MDX) consisted of three parts:
\begin{itemize}
\item Leaderboard A: Label Noise - the labels in the training data were swapped. The organizers provided the audio dataset with strict rules forbidding the usage of external data, including pre-processing and denoising models trained on external data. The challenge was to provide accurate music separation into four stems while using the training data with the noisy labels.
\item Leaderboard B: Bleeding - some stems in the train contained in others, for example, vocals occasionally bleed into drum tracks. The task was to separate them.
\item Leaderboard C: Anything could be used to achieve maximum quality, including pre-trained models and external data.
\end{itemize}

\subsubsection{The Music Demixing Track, Leaderboard C solution}
In the first stage, the vocals track is extracted using the same approach as in the Cinematic sound demixing track, only the models were slightly different. To separate the vocals, we used a combination of three pre-trained models: UVR-MDX1 (checkpoint: Kim\_Vocal\_1.onnx \cite{kimvocal}), UVR-MDX2 (checkpoint: Kim\_Inst.onnx \cite{kiminst}) from Ultimate Vocal Remover project \cite{uvr}, and Demucs\_ft (vocal-only model) from Demucs4 library \cite{demucs4}. The vocals were separated independently by all of these models, then the results were combined with weights. UVR-MDX2 is an instrumental prediction model, so to get the vocals we need to subtract results from the original track. One augmentation technique we used is an inversion. We inverse the mixture by multiplying the waveform vector by -1, mixture$^{-1}$, run the inference on it, and then reverse the results back. Combining inference obtained on the mixture and its inversion provides a test-time augmentation for the waveform.
The models' ensemble is summarized below:
\begin{itemize}
\item $vocals_1$ = UVR-MDX1(mixture, overlap=0.6)
\item $vocals_2$ = orig\_track - UVR-MDX2$^{-1}$(mixture$^{-1}$, overlap=0.6),
\item $vocals_3$ = 0.5 * Demucs4(mixture, 'demucs\_ft', shifts=1, overlap=0.6) + 0.5 * Demucs4$^{-1}$(mixture$^{-1}$, 'demucs\_ft', shifts=1, overlap=0.6)
\end{itemize}

\begin{equation}
vocals = \frac{\sum_{i=1} w_i vocals_i}{\sum_{i=1} w_i}
\end{equation}

We tried different weights for the vocals' ensemble; weights $12$, $8$, and $3$ for UVR-MDX1, UVR-MDX2 and Demucs4, respectively, produced optimal results. 

Then, we obtained the instrumental part by subtracting vocals from the mixture:
\begin{equation}
instr = mixture - vocals
\end{equation}

Next, we applied four different versions of Demucs models with the following settings to the instrumental track only and it's inversion:
\begin{itemize}
\item $bass_1, drums_1, other_1$ = 0.5 * Demucs4(instr, 'demucs\_ft', shifts=1, overlap=0.5) + 0.5 * Demucs4$^{-1}$(instr$^{-1}$, 'demucs\_ft', shifts=1, overlap=0.5) \\
\item $bass_2, drums_2, other_2$ = 0.5 * Demucs4(instr, 'demucs', shifts=1, overlap=0.6) + 0.5 * Demucs4$^{-1}$(instr$^{-1}$, 'demucs', shifts=1, overlap=0.6)\\
\item $bass_3, drums_3, other_3$ = 0.5 * Demucs4(instr, 'demucs\_6s', shifts=1, overlap=0.6) + 0.5 * Demucs4$^{-1}$(instr$^{-1}$, 'demucs\_6s', shifts=1, overlap=0.6)\\
\item $bass_4, drums_4, other_4$ = 0.5 * Demucs3(instr, 'demucs\_mmi', shifts=1, overlap=0.6) + 0.5 * Demucs3$^{-1}$(instr$^{-1}$, 'demucs\_mmi', shifts=1, overlap=0.6)
\end{itemize}

These four models differ in performance, as can be seen in Table~\ref{table:models2} for the Multisong leaderboard. Based on our leaderboards, we combine the models' results with the following weights:
\begin{equation}
\bar{bass} = 19 \cdot bass_1 + 4\cdot bass_2 + 5\cdot bass_3 + 8\cdot bass_4
\end{equation}
\begin{equation}
\bar{drums} = 18\cdot drums_1 + 2\cdot drums_2 + 4\cdot drums_3 + 9\cdot drums_4
\end{equation}
\begin{equation}
\bar{other} = 14\cdot other_1 + 2\cdot other_2 + 5\cdot other_3 + 10\cdot other_4
\end{equation}
Note, the model 'demucs\_mmi' has a slightly different architecture (Demucs3); it can be included with a larger weight for diversification.

Finally, we obtain the values of the final stem tracks as follows:
\begin{equation}
bass = \frac{1}{3}(instr - \bar{other} - \bar{drums} + 2\cdot \bar{bass}) 
\end{equation}
\begin{equation}
drums = \frac{1}{3}(instr - \bar{other} - \bar{bass} + 2\cdot \bar{drums}) 
\end{equation}
\begin{equation}
other = \frac{1}{3}(2\cdot instr - \bar{bass} - \bar{drums} + \bar{other})
\end{equation}

The results for this ensemble are in Table~\ref{tab:final}.
\begin{table}[htbp]
  \centering
  \caption{SDR metrics for the final ensemble on the MultiSong MVSep datasets and MDX23 test sets (leaderboard C).}
  \label{tab:final}
  \begin{tabular}{lccccc}
    \toprule
    Dataset & SDR\_bass & SDR\_drums & SDR\_other & SDR\_vocals & mean\_SDR \\
    \midrule
    MultiSong MVSep& 12.68 & 11.68 & 6.67 & 9.62 & 10.11 \\
    MDX23 public test& 9.87 & 9.52 & 7.43 & 10.81 & 9.41 \\
    MDX23 private test& 9.94 & 9.53 & 7.05 & 10.51 & 9.25 \\
    \bottomrule
  \end{tabular}
\end{table}

The proposed solution achieved the top 3rd result in the challenge.



\section{Conclusions}
This work introduces two new benchmarks for the sound source separation tasks, Synth MVSep and Multisong MVSep. We provide leaderboard based on these datasets and compare popular models for sound demixing and their ensembles. The full leaderboards are dynamic and can be assessed at \url{https://mvsep.com/quality\_checker/}, giving comparisons for a range of models. The benchmark datasets are available for download. 

The current top-performing models for separating vocals from instrumental parts are variations of the MDX algorithm from the open-source project Ultimate Vocal Remover, UVR-MDX~\cite{uvr}.
For the instrumental part (bass, drums, other) separation, Demucs4 HT models variants provide the best results among tested models.  Therefore, ensembles of different models used for vocal and non-vocal stems are expected to provide the top overall performance.

We describe a novel approach for audio separation, based on the ensembling of different models that are suited best for the particular stem. The proposed solution was evaluated in the context of the Sound Demixing Challenge 2023 and achieved top results in different tracks of the competition. The code and the approach are open-sourced on GitHub \cite{zfturbo}. 

Note: The final output will be published after the end of the challenge. 

\begin{ack}
We acknowledge the organizers of the SDX23 challenge for the interesting task, evaluation and the datasets, the authors of all cited open-source libraries for providing models and code. We also want to say big thanks to Bas Curtiz who did the main work for creating MultiSong Dataset and Kimberley Jensen for great vocal models created for UVR project. This research was supported in part through computational resources of HPC facilities at HSE University~\cite{kostenetskiy2021hpc}.
\end{ack}

\medskip

{\small
\bibliographystyle{IEEEtran}
\bibliography{main}
}

\end{document}